\magnification \magstep1
\input amssym.def
\input amssym.tex
\def \sf{\Sigma_F}
\def \si{\Sigma_I}
\def \ep{\epsilon}

\def \ni{\noindent}
\def \rta{\rightarrow}

\def \bs{\bigskip}
\def \in{\indent}
\def \ss{\smallskip}
\def \sclass{S_{\rm class}}
\bigskip
\centerline{\bf Bogoliubov Transformations for Amplitudes in
Black-Hole Evaporation}
\bigskip
\centerline{ A.N.St.J.Farley and P.D.D'Eath }
\smallskip
\bigskip
\centerline{Department of Applied Mathematics and Theoretical Physics,
Centre for Mathematical Sciences,} 
\smallskip
\centerline{University of Cambridge, Wilberforce Road, Cambridge CB3 0WA,
United Kingdom}
\medskip
\centerline{Abstract}
\smallskip
\ni
In a previous Letter, we outlined an approach to the calculation of
quantum amplitudes appropriate for studying the black-hole radiation
which follows gravitational collapse.  This formulation must be
different from the familiar one (which is normally carried out by
considering Bogoliubov transformations), since it yields quantum
amplitudes relating to the final state, and not just the usual
probabilities for outcomes at a late time and large radius.  Our 
approach simply follows Feynman's $+i\ep$ prescription.  Suppose that, 
in specifying the quantum amplitude to be calculated, initial data for
Einstein gravity and (say) a massless scalar field are specified on an
asymptotically-flat space-like hypersurface $\Sigma_I{\,}$, and final
data similarly specified on a hypersurface $\Sigma_F{\,}$, where both
$\Sigma_I$ and $\Sigma_F$ are diffeomorphic to ${\Bbb R}^{3}{\,}$.
Denote by $T$ the (real) Lorentzian proper-time interval between
$\Sigma_I$ and $\Sigma_F{\,}$, as measured at spatial infinity.  Then
rotate: 
$T\rightarrow{\,}{\mid}T{\mid}{\,}\exp(-i\theta){\,},
{\;}0<\theta\leq\pi/2{\,}$.
The {\it classical} boundary-value problem is then expected to become
well-posed on a region of topology $I{\,}\times{\,}{\Bbb R}^{3}{\,}$,
where $I$ is the interval $\bigl[0,{\mid}T{\mid}\bigr]$.  For a locally-
supersymmetric theory, the quantum amplitude is expected to be
dominated by the semi-classical expression $\exp(i S_{\rm class})$,
where $S_{\rm class}$ is the classical action.  Hence, one can find
the Lorentzian quantum amplitude from consideration of the limit 
$\theta\rightarrow 0_{+}{\,}$.  In the usual approach, the only
possible such final surfaces $\Sigma_F$ are in the strong-field region
shortly before the curvature singularity; that is, one cannot have a 
Bogoliubov transformation to a smooth surface 'after the singularity'.
In our complex approach, however, one can put arbitrary smooth
gravitational data on $\Sigma_F{\,}$, provided that it has the correct
mass $M{\,}$; thus we do have Bogoliubov transformations to surfaces
'after the singularity in the Lorentzian-signature geometry' --- the
singularity is simply by-passed in the analytic continuation (see below).  
In this Letter, we consider Bogoliubov transformations in our
approach, and their possible relation to the probability distribution 
and density matrix in the traditional approach.  In particular, we
find that our probability distribution for configurations of the final 
scalar field cannot be expressed in terms of the diagonal elements of some 
density-matrix distribution.\par
\bs
\ni
{\bf 1. Introduction}
\bs
\in
In [1-4], based on the thesis [5], we described a calculation of
the quantum amplitude for an initial spherically-symmetric
configuration of gravity and a massless scalar field $\phi$ to 
make a transition to another nearly-spherically-symmetric
configuration at a very late Lorentzian time $T$, apart from some 
weak perturbations $(h^{(1)}_{ij},\phi^{(1)})_F{\,}$.  The boundary data 
are specified on an initial and a final asymptotically-flat hypersurface 
$\Sigma_{I,F}{\,}$, and the time $T$ is measured at spatial infinity.  
The Lorentzian-signature space-time metric 
${\,}g_{\mu\nu}{\,}(\mu,\nu=0,1,2,3)$ has spatial components 
$h_{ij}=g_{ij}{\,}(i,j=1,2,3)$ on surfaces with 
$t=x^{0}={\rm const.}$, such as $\si$ and $\sf{\,}$. This 
transition could refer to the quantum evaporation of a black hole, 
following gravitational collapse from a slowly-moving, diffuse, 
spherical configuration.  The final configuration could refer to the 
nearly-spherical stream of (massless) radiation, and its corresponding 
Vaidya-like gravitational field [6,7], as viewed on a space-like 
hypersurface $\sf{\,}$, of topology ${\Bbb R}^3$, which cuts through all 
the radiation.  The weak perturbations on $\sf{\,}$, in the present field 
language, correspond to emitted particles.\par
\ss
\in
The calculation proceeds following Feynman's $+i\ep$ prescription [8].  
If $T$ is rotated into the complex: 
$T\rightarrow{\mid}T{\mid}\exp(-i\theta)$, for 
${\,}0<\theta\leq\pi/2{\,}$, then the {\it classical} boundary-value
problem, given the above nearly-spherically-symmetric boundary data, 
is expected to become well-posed [1].\par  
\ss
\in
For a simple example (cf. [10]), consider the 2-dimensional
Euclidean-space Laplace equation
$${{{\partial}^{2}\phi}\over{\partial{\tau}^{2}}}{\;}
+{\;}{{{\partial}^{2}\phi}\over{\partial{x}^{2}}}{\;}
={\;}0{\quad};
{\qquad}{\quad}0<\tau<T{\,},{\quad}-{\infty}<x<+{\infty}{\;},\eqno(1.1)$$
\ni
subject to boundary data of the form (say)
$$\phi(\tau =0,{\,}x){\;}={\;}0{\;},
{\qquad}\phi(\tau =T,{\,}x){\;}={\;}{\phi}_{1}(x){\;},\eqno(1.2)$$
\ni
on the 'initial and final surfaces'.  Here, the boundary data given by
${\phi}_{1}(x)$ should be thought of as a $C^{\infty}$ function, of
rapid decrease as ${\,}{\mid}x{\mid}\rightarrow{\,}\infty{\,}$.\par  
\ss
\in
Let ${\Phi}_{1}(k)$ denote the Fourier transform of
${\phi}_{1}(x){\,}$:
$${\Phi}_{1}(k){\;}={\;}{{1}\over{\sqrt{2\pi}}}{\;}\int_{-\infty}^{\infty}
{\,}e^{-ikx}{\;}{\phi}_{1}(x){\;}dx{\;}.\eqno(1.3)$$
\ni
Then the (unique) solution of the real elliptic Dirichlet
boundary-value problem (1.1,2) above is given by
$${\phi}(\tau,x){\;}={\;}{{1}\over{\sqrt{2\pi}}}{\;}\int_{-\infty}^{\infty}
{\;}e^{ikx}{\;}{{\sinh(k\tau)}\over{\sinh(kT)}}{\;}{\Phi}_{1}(k){\;}dk{\;}.
\eqno(1.4)$$
\ni
In common with any solution of an elliptic partial-differential
equation with analytic coefficients, $\phi$ is automatically a 
(real- or complex-) analytic function of both arguments $\tau$ 
and $x{\,}$ [9].\par
\ss
\in
One can then rotate the Euclidean proper-distance-at-infinity into the
complex: $T{\,}\rightarrow{\,}T\exp(i\theta){\,}$, where 
${\,}0\leq\theta<\pi/2{\,}$.  The Laplace equation (1.1) 
still holds, although it is perhaps more natural to view it with 
respect to coordinates $(y_{\theta},x)$, where 
$y_{\theta}=\tau\exp(i\theta)$, in which case the potential $\phi$ is, 
for each fixed $\theta$, a (complex) solution of the complexified 
Laplace equation
$$e^{2i\theta}{\;}{{{\partial}^{2}\phi}\over{\partial{y_{\theta}}^{2}}}{\;}
+{\;}{{{\partial}^{2}\phi}\over{{\partial}x^{2}}}{\;}={\;}0{\;}.\eqno(1.5)$$
\ni
Provided that ${\,}0\leq\theta <\pi/2{\,}$, this differential 
equation is {\it strongly elliptic} in the language of [11].  We
continue to pose the same boundary data (1.2), with ${\phi}=0$ on the 
initial surface ${\,}y_{\theta}=0{\,}$, and with
$${\phi}(y_{\theta}=T,{\,}x){\;}={\;}{\phi}_{1}(x){\quad}\eqno(1.6)$$
\ni
on the final surface $y_{\theta}=T{\,}$ [equivalently given by
${\tau}=T\exp(-i\theta)$].  The corresponding complexified solution 
$\phi(y_{\theta},x)$ of this boundary-value problem still exists and
is unique, provided always that ${\,}0\leq\theta <\pi/2{\,}$, 
being given by the analytic continuation of Eq.(1.4): 
$$\phi\Bigl({\tau}{\,}={\,}y_{\theta}{\,}\exp(-i\theta),{\,}x\Bigr){\;}
={\;}{{1}\over{\sqrt{2\pi}}}{\;}
\int_{-\infty}^{\infty}{\,}e^{ikx}{\;}
{{\sinh\bigl(k{\,}y_{\theta}\exp(-i\theta)\bigr)}
\over{\sinh\bigl(k{\,}T\exp(-i\theta)\bigr)}}{\quad}{\Phi}_{1}(k){\;}dk{\;}.
\eqno(1.7)$$
\ni
Provided that ${\,}0\leq\theta <\pi/2{\,}$, the denominator 
in the integrand of Eq.(1.7), namely
$$\sinh\bigl(kTe^{-i\theta}\bigr){\;}
={\;}\cos\bigl(kT\sin\theta\bigr){\,}\sinh\bigl(kT\cos\theta\bigr){\,}
-{\,}i{\,}\sin\bigl(kT\sin\theta\bigr){\,}\cosh\bigl(kT\cos\theta\bigr){\;},
\eqno(1.8)$$
\ni
is non-zero for all ${\,}k{\,}{\neq}{\,}0{\,}$.  By examining a small
neighbourhood of $k=0{\,}$, one sees that both numerator and
denominator have simple zeros at $k=0{\,}$, whence the integrand in
Eq.(1.7) is well-behaved for all $k{\,}$.  In this way, it can be
shown that the integral (1.7) provides the (unique) solution of the
boundary-value problem, for ${\,}0\leq\theta <\pi/2{\,}$.\par
\ss
\in
On the other hand, for the exactly Lorentzian case with 
${\theta}=\pi/2{\,}$, one can see that no solution exists for the 
Dirichlet boundary-value problem, for typical choices of the final 
boundary data ${\phi}_{1}(x)$, or equivalently of its Fourier
transform ${\Phi}_{1}(k)$.  This is because, following the
representation (1.4), one now has $\sin(kT)$ for the denominator,
rather than $\sinh(kT)$.  The existence of a (smooth) Lorentzian solution 
$\phi(t,x)$ would then imply that ${\Phi}_{1}(k)$ vanishes for all
those values of $k{\,}$, namely, for all $k=n{\pi}/T$ 
($n$ a positive or negative integer, excluding the case $n=0$), 
such that $\sin(kT)$ vanishes.  A typical ${\phi}_{1}(x)$ will not 
have ${\Phi}_{1}(n{\pi}/T)=0$ for a single value of $n$ 
(integer, ${\neq}{\,}0$).  This argument can be refined, to show
rigorously the non-existence of a solution to the above (non-trivial) 
Dirichlet problem for the flat wave equation.  Thus, in this example, 
the Dirichlet boundary-value problem for the wave equation is badly posed.\par
\ss
\in
In [1] and in this paper, we study the field-theoretic analogue of the
boundary-value problem above, with data specified on an initial and a
final spacelike hypersurface, with either a (Riemannian) proper
distance at spatial infinity, $T$, between the surfaces, or a
complex-rotated separation $T\exp(i\theta)$, where 
$0\leq\theta <\pi/2{\,}$.  For simplicity, we consider Einstein gravity
coupled to a massless scalar field $\phi{\,}$, and first restrict
attention to spherically-symmetric configurations, as giving a large
number of 'background' or 'reference' space-times describing
gravitational collapse to a Schwarzschild black hole in Lorentzian
signature [12,13].  The metric is taken (without loss) in the form
(2.1) below; the corresponding scalar field is of the form
$\phi(t,r)$.  In Riemannian signature, the Einstein field equations are
given in [14].\par
\ss
\in
As yet, very little is known rigorously in the theory of partial
differential equations about existence and uniqueness for this
boundary-value problem -- even for the purely real Riemannian case with
a spherically-symmetric 3-metric $(h_{ij})_{I,F}$ and scalar field
$(\phi)_{I,F}$ specified on the initial and final hypersurfaces, and a
real proper-distance separation $T$ given at spatial infinity.  In the
limit of weak boundary data, and hence, {\it via} the classical
Riemannian field equations, also of weak fields in the interior, one
would expect fixed-point methods, analogous to those used in [15],
to establish existence and uniqueness for the Riemannian
boundary-value problem.  This weak-field Riemannian problem has also
been investigated numerically [14].  For weak scalar boundary data,
global quantities such as the mass $M$ and Euclidean action $I$ appear
to scale quadratically, in accordance with analytic weak-field
estimates [16].\par
\ss
\in
Considering still the real Riemannian case, there does not
'experimentally' appear to be any obstacle to extending the above
(good) results numerically to strong boundary data, except for the
ever-larger amounts of computer time required.  It may be that a
typical pattern will emerge numerically for the general 'shape' of the
classical gravitational and scalar fields, in the limit of strong-field
boundary data.  In that case, it might be possible to find analytic
approximations for the strong-field limit (quite different from those
valid in the weak-field case), which could provide a large amount of
further analytical insight into the solutions of the strong-field
Riemannian boundary-value problem.  It is, of course, such
strong-field boundary-value problems which are of most interest, when
we consider rotating: $T\rightarrow T\exp(i\theta)$, with 
$0\leq\theta <\pi/2{\,}$, towards a Lorentzian time-interval.\par
\ss
\in
The strong-field limit for this boundary-value problem is expected to
correspond, in the Lorentzian limit in which $\theta$ approaches
$\pi/2$ from below (although $\theta$ never attains the value
$\pi/2$), to classical spherically-symmetric Einstein/scalar
solutions which form a singularity, surrounded by a black hole.  As
described in the Abstract above, one expects that arbitrary smooth
gravitational and scalar data can be posed on the final spacelike
hypersurface ${\Sigma}_{F}{\,}$, allowing for a rotation into the complex
of the Riemannian distance-separation $T{\;}(>0)$ between
${\Sigma}_{I}$ and ${\Sigma}_{F}{\,}$, of the form 
$T\rightarrow T\exp(i\theta)$, with ${\,}0\leq\theta <\pi/2{\,}$, and
provided that ${\Sigma}_{F}$ has the correct mass $M$.  The (real)
magnitude $T$ could then be chosen arbitrarily large, corresponding to
a non-singular complex solution of the Einstein/scalar field equations
between ${\Sigma}_{I}$ and a final hypersurface ${\Sigma}_{F}$ at late
time.  In this sense, in the complexified solution, one expects to have
Bogoliubov transformations to surfaces ${\Sigma}_{F}$ which are 'after
the singularity in the Lorentzian-signature geometry', provided always
that one does not take the limit $\theta =\pi/2{\,}$.  In the same
sense, one may say that 'the singularity is simply by-passed in the
analytic continuation'; by analogy with the scalar-field example of 
Eqs.(1.1-8) above, there should be an analytic complexified classical
solution to the boundary-value problem, which only reaches a singular
boundary precisely at Lorentzian signature ${\,}(\theta=\pi/2)$.  Provided
that one retains the freedom to rotate $T$ into the complex (as was
insisted upon also by Feynman [8] in his $+i\epsilon$ prescription),
one has the possibility of circumventing Lorentzian singularities by
deforming time-intervals suitably into the complex, much as one avoids
singularities of functions $f(z)$ in the ordinary complex
$z$-plane.\par
\ss
\in
For the more general Einstein/scalar boundary-value problem, in which
one allows for weak non-spherical perturbations in the boundary data
$(h_{ij},\phi)_{I,F}{\,}$, the (complex) classical action $\sclass$ 
can be studied as a functional of the perturbative final data 
$(h^{(1)}_{ij},\phi^{(1)})_F$ and a function of the complex variable 
$T$, if for simplicity we regard the spherically-symmetric initial 
data $(h_{ij},\phi)_I$ as fixed.  The classical action $\sclass$ can 
be calculated, and thence the semi-classical amplitude, proportional 
to $\exp(i\,\sclass)$.  Finally, the limit $\theta\rightarrow 0_+$ 
can be taken, yielding the Lorentzian amplitude.\par
\ss
\in
This description is clearly very different from that usually adopted
for black-hole evaporation [17-20], which involves the study of
Bogoliubov transformations between an early- and a late-time surface,
where the late-time surface passes inside the black-hole event
horizon, close to the curvature singularity.  The standard approach 
yields probabilities for configurations of outgoing particles, and
a late-time density matrix, but not quantum amplitudes for states
defined typically by weak data over an ${\Bbb R}^{3}$ surface at late
times. as in the present case.  It is intended, in this Letter, 
to make some contact between these two approaches, even though our
approach addresses a wider class of questions than does the usual
approach.  In particular, we show how the Bogoliubov transformation 
between our two ${\Bbb R}^3$-surfaces $\si$ and $\sf$ can be described 
in terms of the quantities appearing in [2,3] (see also [1]).  
The Bogoliubov quantities ${\mid}\beta_{s\omega\ell m}{\mid}^2$ 
for the original calculation of black-hole radiance [17] agree with 
those found by the above boundary-value approach, to high accuracy 
(Sec.3).  In Sec.4, we consider the resulting probability distribution 
for configurations of the perturbative scalar field $\phi^{(1)}$ 
on the final surface $\sf{\,}$.  It is found that this probability 
distribution cannot be expressed in terms of the diagonal elements 
of some density-matrix distribution; thus, the correspondence between
our approach and the density-matrix approach of [18] is limited.\par
\bs
\in
{\bf 2. Bogoliubov transformations between $R^{3}$ surfaces}
\bs
\in
As in [1-4], we write the Lorentzian-signature classical 'background'
metric in the form
$$ds^{2}{\;}={\;}-{\,}e^{b(t,r)}dt^{2}{\,}+{\,}e^{a(t,r)}dr^{2}+
{\,}r^{2}(d{\theta}^{2}{\,}+{\,}{\sin}^{2}\theta{\,}d{\phi}^{2}){\;}.
\eqno(2.1)$$
\noindent
For later reference, we also define the 'mass function' $m(t,r)$ by
$$\exp\Bigl(-a(t,r)\Bigr){\;}
={\;}1{\,}-{{2m(t,r)}\over{r}}{\;}.\eqno(2.2)$$
\noindent
In the boundary-value problem outlined above, let us write
$(\gamma_{\mu\nu}{\,}, \Phi)$ for the 'background' spherically-symmetric
metric and scalar field.  Writing also $\nabla_{\mu}$ for the
background covariant derivative, consider a linearised classical
solution $\phi^{(1)}$ of the scalar wave equation
$$\nabla^{\mu}\nabla_{\mu}\phi^{(1)}{\;}={\;}0{\;},\eqno(2.3)$$
\ni
Consider a basis $\{f_{\omega'\ell m}(x)\}$ of (nearly-) separated
mode solutions adapted to $\si$ and another such basis
$\{p_{\omega\ell m}(x)\}$ adapted to $\sf{\,}$.  On $\si$, the 
$\{f_{\omega'\ell m}\}$ are chosen to be an orthonormal, complete
family of complex solutions of the wave equations, which contain only
positive frequencies $(\omega'>0)$.  Correspondingly for the 
$\{p_{\omega\ell m}\}$ on $\sf$.  In a standard fashion [19], one
expands out $\phi^{(1)}(x)$ in terms of one or the other basis, as
$$\eqalignno{\phi^{(1)}(x){\;}&={\;}\sum^{\infty}_{\ell =0}{\;}{\;}
\sum^{\ell}_{m=-\ell}{\;}\int^{\infty}_{0}{\;}d\omega^{\prime}{\;}
\Bigl[c_{\omega^{\prime}\ell m}{\;}f_{\omega^{\prime}\ell m}(x){\;} 
+{\;}{\rm c.c.}\Bigr]{\;},&(2.4)\cr
\phi^{(1)}(x){\;}&={\;}\sum^{\infty}_{\ell =0}{\;}{\;}
\sum^{\ell}_{m =-\ell}{\;}\int^{\infty}_{0}{\;}d\omega{\;}
\Bigl[b_{\omega\ell m}{\;}p_{\omega \ell m}(x){\;} 
+{\;}{\rm c.c.}\Bigr]{\;}.&(2.5)\cr}$$
\ni
Note that we shall assume throughout that the asymptotic time interval
${\,}T={\mid}T{\mid}{\,}\exp(-i\theta)$ has negative imaginary part.
For convenience of visualisation, one might wish to regard the angle
$\theta$ as being extremely small.  Nevertheless, one only takes
$\theta \rightarrow 0_+$ at the end of the calculation.  Thus, this
Section deals with a larger class of Bogoliubov transformations than
those appearing in the usual formulation.\par
\ss
\in
The bases above may be chosen to have the form 
$$\phi_{\omega\ell m}(x){\;}={\;}N(\omega){\;}
{{R_{\omega\ell}(r)}\over{r}}{\;} 
e^{-i\omega t}{\;}Y_{\ell m}(\Omega){\;},\eqno(2.6)$$
\ni
where $N(\omega)$ is a normalisation factor.  On $\si{\,}$, we require
${\,}\phi_{\omega \ell m}{\,}\propto{\,}e^{-i\omega v}{\,}$ at large
$r$, where $v=t+r^{*}$ and where ${\,}r^{*}{\,}\sim{\,}r^{*}_{s}{\;}$,
with    
$$r^{*}_{s}{\;}={\;}r{\,}+{\,}2M{\,}
\ln\Bigl(\bigl(r/2M\bigr)-1\Bigr){\;}\eqno(2.7)$$ 
\ss
\ni
denoting the Regge-Wheeler radial coordinate [21,22] in an exactly
Schwarzschild geometry of mass $M$.  On $\sf{\,}$, we require
${\,}\phi_{\omega\ell m}{\,}\propto{\,}e^{-i\omega u}{\,}$ at large $r$, where
${\,}u=t-r^*{\,}$.  Thus, the $\{f_{\omega'\ell m}\}$ are purely ingoing
at infinity, while the $\{p_{\omega\ell m}\}$ are purely outgoing.\par
\ss
\in
Since the $\{f_{\omega'\ell m}\}$ give a complete orthonormal set on
$\si{\,}$, a typical basis function (solution) $p_{\omega \ell m}$ on
$\si$ may be expanded as
$$p_{\omega\ell m}{\;}={\;}\int^{\infty}_{0}d\omega'{\;}
\Bigl(\alpha_{\omega'\omega\ell m}{\;}f_{\omega'\ell m}{\;}
+{\;}\beta_{\omega'\omega\ell m}{\;}f^{*}_{\omega'\ell, -m}\Bigr){\;}
,\eqno(2.8)$$
\ni
The sets of complex numbers $\{\alpha_{\omega'\omega \ell m}\}$ and 
$\{\beta_{\omega'\omega\ell m}\}$ are the Bogoliubov coefficients, and
the standard properties of Bogoliubov transformations are, for
example, given in [13].\par
\ss
\in
In our treatment [1-5] of linearised scalar perturbations
$\phi^{(1)}(x)$, propagating at late times through the Vaidya-like
region containing outgoing black-hole radiation, almost all modes are
adiabatic.  That is, a general solution of Eq.(2.1) in this region can
be built up from separated solutions of the form (2.6), with 
$\omega=-k$ and ${\,}R_{\omega\ell}(r){\,}\sim{\,}\xi_{k\ell}(r){\,}$.  
Here, $\xi_{k\ell}(r)$ obeys 
$${{\partial^{2}\xi_{k\ell}}\over {\partial{\,}r^{*2}}}{\;}
+{\;}\bigl(k^{2}-V_{\ell}\bigr){\;}\xi_{k\ell}{\;}={\;}0{\;},\eqno(2.9)$$
\ni
where, as usual,
$${\partial\over {\partial r^{*}}}{\;}
={\;}e^{(b-a)/2}{\;}{{\partial\over{\partial r}}}{\quad},{\qquad}{\qquad} 
e^{-a}{\;}\simeq{\;}e^{b}{\;}\simeq{\;}1-{{2m(t,r)}\over{r}}{\;},
\eqno(2.10)$$
\noindent
and
$$V_{\ell}(t,r){\;}={\;}{{e^{b(t,r)}}\over{r^{2}}}\biggl(\ell(\ell +1)
+{{2m(t,r)}\over{r}}\biggr).\eqno(2.11)$$
The 'left' boundary condition on ${\,}\xi_{k\ell}{\,}$ 
as ${\,}r\rightarrow 0{\,}$ is
$$\xi_{k\ell}(0){\;}={\;}0{\;}.\eqno(2.12)$$
\ni
The 'right' boundary condition, as ${\,}r\rightarrow\infty{\,}$, is
$$\xi_{k\ell}(r)\quad\sim\quad
z_{k\ell}{\;}\exp\bigl(i{\,}k{\,}r^{*}_{s}\bigr){\;}
+{\;}z^{*}_{k\ell}{\;}\exp\bigl(-i{\,}k{\,}r^{*}_{s}\bigr){\;},\eqno(2.13)$$
\noindent
where, for each $(k,\ell)$, ${\,}z_{k\ell}$ is a dimensionless complex
constant.  We then wrote out $\phi^{(1)}$ in terms of 'coordinates'
$\{a_{k\ell m}\}$ on the final surface $\sf{\,}$, 
$$\phi^{(1)}(x)\Bigl\arrowvert_{\Sigma_F}{\quad} 
={\quad}{1 \over r}{\;}{\,}\sum^{\infty}_{\ell = 0}{\;}
\sum^{\ell}_{m=-\ell}{\;}\int^{\infty}_{-\infty}{\;}
dk{\;}{\,}a_{k\ell m}{\;}\xi_{k\ell}(r){\;}Y_{\ell m}(\Omega){\;}.
\eqno(2.14)$$
\ss
\in
In the nearly-Lorentzian case, the late-time behaviour of quantum
amplitudes [1-5] will be dominated by the eigen-frequencies
$$k_{n}{\;}={\;}{{n\,\pi}\over{T}}{\;}.\eqno(2.15)$$
\noindent
Correspondingly, one discretises the frequency integral (2.11).  It
will also be convenient to define real functions $\{f_{n\ell}(r)\}$
for ${\,}n=1,2,\ldots;{\quad}\ell =0,1,2,\ldots$, such that
$$\xi_{n\ell}(r){\;}
={\;}\Bigl(z_{n\ell}{\;}e^{ik_{n}r^{*}}
+{\;}z^{*}_{n\ell}{\;}e^{-ik_{n}r^{*}}\Bigr){\;}f_{n\ell}(r){\;},\eqno(2.16)$$
\noindent
where ${\,}f_{n\ell}(r)\rightarrow 1{\;}$ 
as ${\;}r\rightarrow\infty{\;},\quad f_{n\ell}(r)\rightarrow 0{\;}$ 
as ${\;}r\rightarrow 0{\;}$.\par
\ss
\in
It is now possible to make a comparison of this treatment of the quantum
amplitude from initial data on ${\Bbb R}^3$ to final data on 
${\Bbb R}^3$, with the corresponding description in terms of
Bogoliubov transformations (again, in our formulation, not in the
standard formulation).  One takes a discretised version of Eq.(2.5), 
appropriate to the final surface $\sf{\,}$, and compares the expansions 
of $\phi^{(1)}$ in basis functions on $\sf{\,}$.  In particular, one
finds that
$${\mid}b_{n\ell m}{\mid}^{2}{\quad}
={\quad}2{\,}\pi{\,}k_{n}{\,}{\mid}z_{n\ell}{\mid}^{2}{\;}
{\mid}a_{n\ell m} 
+a_{-n\ell m}{\mid}^{2}{\quad}
={\quad}2{\,}f_{\ell m}(k_{n}){\;},\eqno(2.17)$$
\ni
where $\{b_{n\ell m}\}$ are the coefficients appearing in Eq.(2.5).
We also find an expression for ${\mid}b_{\omega\ell m}{\mid}^{2}$ in
terms of Bogoliubov coefficients.  Including an implicit summation
over the index $m$:
$$\eqalign{{\mid}b_{\omega\ell m}{\mid}^{2}{\;}&={\;}\int^{\infty}_{0}
d\omega'{\,}\int^{\infty}_{0}
d\omega''{\,}{\,}\biggl(\alpha^{*}_{\omega'\omega\ell}{\;}
\alpha_{\omega''\omega\ell}{\;}    
+{\;}\beta^{*}_{\omega''\omega\ell}{\;}\beta_{\omega'\omega\ell}\biggr)
{\;}c_{\omega'\ell m}{\;}c^{*}_{\omega''\ell m}\cr
&-2\int^{\infty}_{0}d\omega'\int^{\infty}_{0}d\omega''
{\;}{\rm Re}\biggl(\beta_{\omega''\omega\ell}{\;}
\alpha^{*}_{\omega'\omega\ell}{\;} 
c_{\omega'\ell m}{\;}c_{\omega''\ell,-m}\biggr){\;}.\cr}\eqno(2.18)$$
\ni
{\bf 3. Particle emission rates for a final $R^{3}$ surface}
\bs
\in
The total energy of the late-time Einstein/scalar system must be equal
to $M$, the ADM mass [22] or energy of the original configuration.
The 'second-variation' contribution $E^{(2)}$ to $M$, namely the
contribution from all the particles produced, is
$$E^{(2)}{\;}={\;}\sum_{\ell m}{\;}\int^{\infty}_{0}{\;}d\omega{\;}\omega{\;} 
{\mid}b_{\omega\ell m}{\mid}^{2}{\;}.\eqno(3.1)$$
\noindent
This can also be written in terms of the 'harmonic-oscillator
coordinates' ${\quad}\{c_{\omega\ell m}\}{\quad}$ of Eq.(2.4), 
on using Eq.(2.18).  Further, the expected number 
$<n_{\omega \ell m}>$ of particles emitted [5] is
$$<n_{\omega\ell m}>{\;}{\,}
={\;}{\,}{\mid}\beta_{\omega\ell}{\mid}^{2}{\;},\eqno(3.2)$$ 
\noindent
where ${\mid}\beta_{\omega\ell}{\mid}^{2}$ is defined by
$$\int^{\infty}_{0}d\omega'{\;}\beta_{\omega'\omega\ell}{\;}
\beta^{*}_{\omega'\omega''\ell}{\;}{\,}
={\;}{\,}{\mid}\beta_{\omega\ell}{\mid}^{2}{\;}\delta(\omega,\omega'')
{\;}.\eqno(3.3)$$
\ss
\in
In the original, standard, calculation of Bogoliubov coefficients and of
probabilities for particle emission by a non-rotating black 
hole [17], neglecting back-reaction, it was found that 
$${\mid}\beta_{s\omega\ell m}{\mid}^{2}{\;}{\,}
={\;}{\,}\Gamma_{s\omega\ell m}(\tilde m){\;}
\biggl(e^{4\pi{\tilde m}}-(-1)^{2s}\biggr)^{-1}{\;},\eqno(3.4)$$
\noindent
where $\Gamma_{s\omega\ell m}(\tilde m)$ is the transmission
probability over the centrifugal barrier [23] and 
$\tilde m=2M\omega{\,}$.  This calculation, of course, referred to the 
case where one did not have a final surface $\sf$ of topology 
${\Bbb R}^{3}$, with a freely-chosen metric on it, subject to the mass 
being $M$.  But, because of the very-high-frequency (adiabatic) method 
in which this expression for ${\mid}\beta_{s\omega\ell m}{\mid}^{2}$ 
was calculated, it should still be valid (up to tiny corrections) in 
our present ${\Bbb R}^{3}$ case, noting that the calculation of 
scattering over the centrifugal barrier should allow for the extremely 
slow time- and radial-dependence of the mass function $m(t,r)$ 
in Eq.(2.9-11).\par
\bs
\ni
{\bf 4. Probabilities and density matrices}
\bs
\in
Following from the Lorentzian quantum amplitude calculated in [1-5],
for perturbative spin-$0$ data given on the final surface $\sf{\,}$, while
taking (for simplicity) exactly spherically-symmetric 'background' 
data on $\si{\,}$, one finds the corresponding conditional probability 
density over the final perturbative scalar boundary data 
${\,}\phi^{(1)}{\mid}_{\sf}{\,}$, given that 
${\,}\phi^{(1)}{\mid}_{\si}{\,}=0{\,}$.  
For a large (real) Lorentzian time-interval $T$, this is  
$$P\Bigl[\{a_{k\ell m}\};{\mid}T{\mid}\Bigr]{\;}{\,}
={\;}{\,}{\hat N}{\;}e^{-\delta{\mid}T{\mid}M_{I}}{\;}
\exp\biggl(-2{\,}{\rm Im}{\,}S^{(2)}_{\rm class}\Bigl[\{a_{k\ell m}\};
{\mid}T{\mid}\Bigr]\biggr){\;},\eqno(4.1)$$
\noindent
where $\hat N$ is a suitable normalisation factor.  The quantity
$S^{(2)}_{\rm class}{\,}\Bigl[\{a_{k\ell m}\};{\,}T\Bigr]$ is calculated
in [3].\par
\ss
\in
The probability distribution (4.1) arises, of course, from squaring a
quantum amplitude.  But, given the thermal nature of black-hole
evaporation in the usual description, one can still ask whether the
probability distribution can be viewed in terms of the diagonal
elements of some density matrix in a suitable basis [18-20].  This is
simpler to understand if one assumes the relations
$$\int^{\infty}_{0}d\omega{\;}\beta^{*}_{\omega''\omega\ell}{\;}
\beta_{\omega'\omega\ell}{\;}{\,}
={\;}{\,}{\mid}\beta_{\omega'\ell}{\mid}^{2}
{\;}{\,}\delta(\omega',\omega''){\;},\eqno(4.2)$$
$$\int^{\infty}_{0}d\omega{\;}\beta_{\omega''\omega\ell}{\;}
\alpha^{*}_{\omega'\omega\ell}{\;}{\,}={\;}{\,}0{\;}.\eqno(4.3)$$
\noindent
These conditions (4.2,3) do hold for the steady-state Bogoliubov
coefficients in the calculation which neglects back-reaction on the
metric [17].  One needs to check whether this diagonal form persists
under adiabatic propagation through a slowly-varying potential.  The
slow variation of the potential with $t$ and $r$ occurs, as above, 
because, in the Vaidya metric [6] describing the radiative part of 
the space-time, the Schwarzschild mass $M$ is replaced by the mass 
function $m(t,r)$ of Eq.(2.2,10).\par
\ss
\in
We shall need the relation [19]
$${\mid}\alpha_{\omega'\ell}{\mid}^{2}{\,} 
-{\,}{\mid}\beta_{\omega'\ell}{\mid}^{2}{\;}{\,}
={\;}1{\,}{\;}.\eqno(4.4)$$
\noindent
From these equations, one finds
$$P\biggl[\{a_{\omega\ell m}\}\biggr]{\;}
={\;}P\biggl[\{c_{\omega\ell m}\}\biggr]
={\;}\hat N{\;}\prod_{n\ell m}
\exp\Biggl[{{-2\pi}\over{T}}
\biggl(1+2{\,}{\mid}\beta_{n\ell}{\mid}^{2}\biggl){\;}
{\mid}c_{n\ell m}{\mid}^{2}\Biggr]{\;}.\eqno(4.5)$$
\ni
As expected in the linearised theory, the modes evolve independently,
corresponding to the product over $n\ell m$.\par
\ss
\in
In studying the question of whether 
$P\Bigl[\{c_{\omega\ell m}\}\Bigr]$ can be related to a 
density-matrix distribution, we shall need the relation
$${{1}\over{(1-s)}}\exp\Bigl[-xs/(1-s)\Bigr]{\quad}
={\quad}\sum^{\infty}_{k=0}{\;}L_{k}(x){\;}s^{k},
{\qquad}{\mid}s{\mid}{\;}<{\;}1{\;},\eqno(4.6)$$ 
\noindent
which gives the generating function for the Laguerre polynomials
$\{L_{k}(x)\}$ [24].  We set
$$s{\;}
={\;}\Bigl{\arrowvert}
{{\beta_{n\ell}}\over{\alpha_{n\ell}}}\Bigl{\arrowvert}^{2}{\;}<{\;}1
\eqno(4.7)$$
\noindent
and 
$$x{\;}{\,}
={\;}{\,}2{\;}X_{n\ell m}{\;}{\,}
={\;}{\,}4{\;}(\Delta\omega_{n}){\;}{\mid}c_{n\ell m}{\mid}^{2}{\;},
\eqno(4.8)$$
\noindent
which is dimensionless.  Writing $j = n\ell m$, we find
$$\eqalign{P[\{c_{j}\}]{\;}&={\;}\hat N{\;}\prod_{j}{\;}
\exp(-X_{j}){\;}\exp\Bigl(-{\,}2{\,}{\mid}\beta_{j}{\mid}^{2} X_{j}\Bigr)\cr
&={\;}\hat N{\;}\prod_{j}{\;}\exp(-X_{j}){\;}\sum^{\infty}_{k_{j}=0}
P(k_{j}){\;}L_{k_j}(2X_{j})\cr
&={\;}\hat N{\;}\sum_{k_{j}}{\;}P\Bigl(k_{j}\Bigr){\;}
\prod_{j}{\;}\exp(-X_{j}){\;}L_{k_{j}}(2X_{j}){\;},\cr}\eqno(4.9)$$
\noindent
where $P(k_{j})$ is defined as
$$P(k_{j}){\;}
={\;}{{1}\over{{\mid}\alpha_{j}{\mid}^{2}}}{\;}
\Bigl{\arrowvert}{{\beta_{j}}\over{\alpha_{j}}}
\Bigr{\arrowvert}^{2k_{n\ell m}}{\;},\eqno(4.10)$$
\noindent   
giving the (normalised) probability to observe the field in the state
$k_j$.  Equivalently, $P(k_{j})$ is the probability of finding $k_j$
particles outgoing at future null infinity, in the mode $(n\ell m)$.
Eq.(4.9) describes the probability distribution $P[\{c_{j}\}]$ for the
final scalar field, with the help of the Bose-Einstein
distribution (4.10).  From Eq.(4.5), $P[\{c_{j}\}]$ depends only on
the modulus ${\mid}c_{j}{\mid}$ of the complex number $c_j{\,}$;
hence, the sum in Eq.(4.9) is diagonal.\par
\ss
\in
In Eq.(4.9), the Laguerre polynomial $L_{k}(x)$ has $k$ real, distinct
roots.  For positive argument $X_{j}>0{\,}$ and excited states
$(k_{j}>0)$, the function $\exp(-X_{j}){\;}L_{k_{j}}(2X_{j})$ takes
negative values for certain ranges of $X_j{\,}$.  Therefore, this function
cannot be interpreted as a probability density.\par
\ss
\in
For comparison, consider a density matrix of the form
$$\rho{\,}(x,y){\;}
={\;}\sum_{i}{\;}w_{i}{\;}\psi_{i}(x){\;}\psi^{*}_{i}(y){\;},\eqno(4.11)$$
\noindent
where the $\{\psi_{i}(x)\}$ form a complete set and $w_i$ is the
probability to be in the state $\psi_{i}(x)$.  If the system is in the
single state defined by $\psi_{i}(x)$, then the probability density
for observing $x{\,}$ is ${\,}{\mid}\psi_{i}(x){\mid}^{2}{\,}.$  
There is not a probabilistic interpretation for the full 
${\,}\rho(x,y){\,}$, but the diagonal contribution 
$$P(x){\;}
={\;}\rho(x,x){\;}
={\;}\sum_{i}{\;}w_{i}{\;}{\mid}\psi_{i}(x){\mid}^{2}\eqno(4.12)$$
\noindent
is the probability density for observing the coordinate $x{\,}$.  In our
example, if the field is in the $k_j$-th state, then 
$\hat N{\,}\exp(-X_{j}){\;}L_{k_{j}}(2X_{j})$ is the 'probability density' 
for the 'harmonic oscillator coordinate' $X_j{\,}$.  We then obtain the
statistical 'probability' distribution (4.9) on multiplying by the
probability $P(k_{j})$ to be in the $k_j$-th state, then summing over
all possible states $k_j$.  This gives an interesting way of regarding
the probability distribution (4.9), but it does not provide a sensible
density matrix, based on the quantum amplitude 
$\exp\Bigl(i{\,}S^{(2)}_{\rm {class}}{\,}\bigl[\{a_{k\ell m}\}, 
T\bigr]\Bigr)$ of [3-5].  This shows that the density-matrix approach, 
based conceptually on linearised quantum field theory around a
classical space-time, does not apparently replicate the results of
a fully quantum approach (at least, not in the locally-supersymmetric
case). \par
\bigskip
\noindent
{\bf 5. Conclusion}
\bs
\in
Here, we have made some connection between the boundary-value approach
of [1-5] to calculating quantum amplitudes for generic late-time
configurations defined (say) on ${\Bbb R^{3}}$, which involves a rotation of
the time-at-infinity: $T\rta{\mid}T{\mid}{\,}\exp(-i\theta)$ into the
complex, and the familiar Bogoliubov-coefficient approach to
calculating probabilities in black-hole evaporation, with a final
hypersurface near the space-time singularity.  In the case that
both the initial and the final space-like hypersurfaces $\si,{\,}\sf$ have
the topology of ${\Bbb R}^3$, there is a relation between the
Bogoliubov transformations in the two different approaches. 
The ${\mid}\beta_{s\omega \ell m}{\mid}^2$
quantities for the original (Bogoliubov) calculation of black-hole
radiance and for the topologically-${\Bbb R}^3$ approach agree, at
least very closely.  However, the probability distribution calculated
from the '${\Bbb R}^{3}$' quantum amplitude cannot be expressed in
terms of the diagonal elements of some density-matrix distribution, 
so supporting the view that the two approaches are genuinely different.\par
\bs
\ni
{\bf Acknowledgements}\par
\bs
\in
We are very grateful to the referee, first, for suggestions which helped
greatly to clarify the distinction between our complex-time approach
and the traditional approach, and, second, for other more detailed
comments.\par  
\bs
\parindent = 1 pt
{\bf References}
\bs
\in
[1] A.N.St.J.Farley and P.D.D'Eath, Phys Lett. B {\bf 601}
(2004) 184. \par          
\indent
[2] A.N.St.J.Farley and P.D.D'Eath, 'Quantum Amplitudes in Black-Hole 
Evaporation; I. Complex Approach', submitted for publication (2005).\par
\in
[3] A.N.St.J.Farley and P.D.D'Eath, 'Quantum Amplitudes in Black-Hole 
Evaporation: II. Spin-0 Amplitude', submitted for publication (2005).\par
\in
[4] A.N.St.J.Farley and P.D.D'Eath, 'Bogoliubov Transformations 
in Black-Hole Evaporation', submitted for publication (2005).\par
\in
[5] A.N.St.J.Farley, 'Quantum Amplitudes in Black-Hole 
Evaporation', Cambridge Ph.D. dissertation, approved 2002 (unpublished).\par
\in
[6] P.C.Vaidya, Proc. Indian Acad. Sci. {\bf A33} (1951) 264.\par 
\indent
[7] A.N.St.J.Farley and P.D.D'Eath, 'Vaidya Space-Time and Black-Hole 
Evaporation', submitted for publication (2005).\par
\in
[8] R.P.Feynman and A.R.Hibbs, Quantum Mechanics and Path Integrals, 
McGraw-Hill, New York, 1965.\par
\indent
[9] P.R.Garabedian, Partial Differential Equations, Wiley, New York, 1964.\par
\indent 
[10] P.D.D'Eath, Supersymmetric Quantum Cosmology, Cambridge
Univ. Press, Cambridge, 1996.\par
\indent
[11] W.McLean, Strongly Elliptic Systems and Boundary Integral
Equations, Cambridge Univ. Press, Cambridge, 2000.\par
\indent
[12] D.Christodoulou, Commun. Math. Phys. {\bf 105} (1986) 337;
{\bf 106} (1986) 587; {\bf 109} (1987) 591; {\bf 109} (1987) 613.\par
\indent
[13] M.W.Choptuik, Critical behaviour in massless scalar field
collapse, in R.d'Inverno (Ed.), Approaches to Numerical Relativity,
Cambridge Univ. Press, Cambridge, 1992.\par 
\indent
[14] P.D.D'Eath, A.Sornborger, Class. Quantum Grav. {\bf 15} (1998) 3435.\par
\indent
[15] P.D.D'Eath, Ann. Phys. (N.Y.), {\bf 98} (1976) 237.\par
\indent
[16] P.D.D'Eath, in preparation.\par
\indent
[17] S.W.Hawking,  Commun. Math. Phys. {\bf 43} (1975) 199.\par
\indent
[18] S.W.Hawking,  Phys. Rev. D {\bf 14} (1976) 2460.\par
\indent
[19] N.D.Birrell and P.C.W.Davies, Quantum fields in curved space,
Cambridge Univ. Press, Cambridge, 1982.\par
\indent
[20] V.P.Frolov and I.D.Novikov, Black Hole Physics, Kluwer Academic,
Dordrecht, 1998.\par
\indent
[21] T.Regge and J.A.Wheeler, Phys. Rev. {\bf 108} (1957) 1063.\par 
\indent
[22] C.W.Misner, K.S.Thorne and J.A.Wheeler, Gravitation, Freeman, 
San Francisco, 1973.\par
\indent
[23] J.A.H.Futterman, F.A.Handler and R.A.Matzner, Scattering from
Black Holes, Cambridge Univ. Press, Cambridge, 1988.\par
\indent
[24] I.S.Gradshteyn and I.M.Ryzhik, Tables of Integrals, Series and 
Products, Academic Press, New York, 1965.\par 

\end